\documentclass[12pt]{article}

\usepackage[fleqn]{amsmath}
\usepackage{amsfonts}
\usepackage{amssymb}

\usepackage{epsfig}
\usepackage{geometry}
\usepackage{ctable}

\newcommand{\cC}{{\mathcal{C}}}

\newcommand{\muV}{{\boldsymbol{\mu}}}

\newcommand{\rV}{{\boldsymbol{r}}}

\newcommand{\Nset}{{\mathbb{N}}}
\newcommand{\Rset}{{\mathbb{R}}}
\newcommand{\Sset}{{\mathbb{S}}}

\begin{document}
\title{Testing Lennard-Jones Clusters for Optimality}

\author{\sc{Michael K.-H. Kiessling}$^*$\\
    {\small{Department of Mathematics}}\\
    {\small{Rutgers, The State University of New Jersey}}\\
     {\small{110 Frelinghuysen Rd., Piscataway, NJ 08854}}\\
     {\small{United States of America}}\\
     {\small{email: miki@math.rutgers.edu}} }

\date{Revised version of June 07, 2023\vspace{-1truecm}} 

\maketitle

\begin{abstract}\noindent
This note advertises a simple necessary condition for optimality that any list $N\mapsto v^x(N)$ 
of computer-generated putative lowest average pair energies $v^x(N)$ of clusters that consist of $N$
monomers has to satisfy, whenever the monomers interact with each other through pair forces satisfying
Newton's ``actio equals re-actio.''
 These can be quite complicated, as for instance in the TIP5P model with five-site potential for a 
rigid tetrahedral-shaped H$_2$O monomer of water, or as simple as the Lennard-Jones single-site potential
for the center of an atomic monomer (which is also used for one site of the H$_2$O monomer in the TIP5P model,
that in addition has four peripheral sites with Coulomb potentials).
 The empirical usefulness of the necessary condition is demonstrated by testing a list of
publicly available Lennard-Jones cluster data that have been pooled from 17 sources,
covering the interval $2\leq N\leq 1610$ without gaps.
 The data point for $N=447$ failed this test, meaning the listed $447$-particle Lennard-Jones cluster 
energy was not optimal.
 To implement this test for optimality in search algorithms for putatively optimal configurations is an easy task. 
Publishing only the data that pass the test would increase the odds that these are actually optimal, without guaranteeing
it, though.
\end{abstract}

\vfill

\hrule
\smallskip
$^*$ {\footnotesize{Corresponding Author.}}

\smallskip
\copyright {\footnotesize{(2023) The author. Reproduction, in its entirety, for non-commercial purposes is permitted.}}

\section{Introduction}\vspace{-.3truecm}
 Increasingly more powerful computer hard- and software has lead to a bonanza of 
data with putative potential-energy ground states of $N$-body clusters for various types of pair interactions of interest to 
theoretical chemistry and physics.
 Such pair interactions include the familiar ones that go by the names of Coulomb, Lennard-Jones, Morse, and others, 
which are quite simple pair interactions that depend only on the distance between the centers of two atoms. 
 Yet they also include more complicated and less widely known ones such as the $n$-site TIPnP interactions 
between two rigid tetrahedral H$_2$O monomers in models of water; see e.g. \cite{MahoJorg} for the $n=5$ model.
 Other examples will be mentioned further below. 
 A wealth of data is available at the Cambridge Cluster Energy Landscape website \cite{CCD}, 
which has a link to the Nankai University's Chemoinformatics webpage \cite{Shao} with further cluster data.
 The qualifier ``putative'' is meant to remind the reader that the vast majority of data is not known for sure to be optimal,
because this optimization problem belongs to the class of NP-hard problems \cite{WilleVennik}, \cite{Adip}.
 As such it is of interest also in computer and complexity science.

 Since there may not exist a deterministic algorithm that finds optimal large $N$-monomer clusters in polynomial time, 
most of the data are ``computer-experimental'' data.
 In this situation it is desirable to have necessary conditions for optimality that can be used to test lists of 
putatively optimal data of clusters with pair interactions.

 A few years ago the author inspected various lists of publicly available putative ground state energy data for 
the so-called Thomson problem and some of its generalizations, subjecting those data lists to a simple 
optimality criterion that compares $N$-particle with $N+1$-particle clusters; see \cite{KieJSP}.
	 That study revealed about two dozen non-optimal energies, mostly for $N$ bigger than 1800, yet one list 
	had two non-optimal data also for $N\approx 100$.
 In a subsequent work \cite{NBK} three non-optimal data with $N\approx 200$ were identified in a list 
of data points for $2\leq N\leq 200$ that had passed the optimality test used in \cite{KieJSP}.
 In \cite{NBK} a different, though only conjectured optimality condition was used that has yet to be proved rigorously.
 After finding three data points that failed the conjectured optimality condition, the authors of \cite{NBK} did seek --- 
and did find --- lower-energy configurations with pertinent $N$, and after replacing the test-failing data points with the 
data of the lower-energy configurations the conjectured optimality condition became satisfied.

 The present paper is meant to advertise the necessary condition for optimality of \cite{KieJSP} in a more appealing manner, 
and for a much larger class of cluster models than envisioned in \cite{KieJSP}; namely, the necessary condition applies to 
all models of monomer clusters whose total energy is a sum of (unordered) pair energies, no matter how complicated. 
This includes cluster models of rigid-shaped monomers with multi-site interactions, such as the already mentioned water cluster models,
and also a family of cluster models of a rigid rod-shaped bacteriophage \cite{SOKMW}, 
as well as some models of viral capsids as clusters of protein monomers \cite{FCW}, the latter two with coarse-grained
pair interactions. 
 These more complicated monomer cluster models were brought to the attention of the author by David Wales, respectively 
by the referee.

 To demonstrate the power of the criterion, we follow up on our remark in section 4 of \cite{KieJSP} and
apply it to the benchmark problem of so-called Lennard-Jones clusters, see e.g.
\cite{Northby},
\cite{WD},
\cite{Leary},
\cite{DoyeMillerWales},
\cite{Xue},
\cite{LocatelliSchoen},
\cite{LXH},
\cite{Mueller},
\cite{Cameron},
\cite{CameronVEijnden},
\cite{FormanCameron},
\cite{BRa},
\cite{BRb},
\cite{YuETal},
\cite{CCD},
\cite{Doye},
\cite{Shao},
and references therein.
 At the website \cite{CCD} one finds a link to \cite{Doye} that lists putatively optimal data pooled from 17 different 
stated sources, and a link to one of these sources: the website \cite{Shao} where one finds some of these and many other data. 
 If poetical licence is permitted, the list features the best-of-the-best of putatively lowest energy data for Lennard-Jones clusters.
 We have collected all these data into a single long list with $2\leq N\leq 1610$ and subjected it to the test criterion of \cite{KieJSP}.
 Not all data passed the test; we rest our case. 

 Since the test criterion is based on a necessary but not sufficient condition for optimality, subjecting 
empirical lists of data to it can only find non-optimal data; it cannot in itself confirm an optimal answer.
Yet the criterion also yields an estimate of the error made whenever a data point fails the test, 
by providing an upper bound on the energy of the actual optimizer.
In the same vein, it provides an upper bound on the energy of any missing optimal configuration in a data gap ---
for instance, suppose one has computed putatively lowest energy configurations for any $N$-body cluster with $N$ even, 
i.e. $N=2K$ with $K\in\Nset$; 
then our necessary condition for optimality yields an upper bound on the optimal energy for each $N$-body cluster with $N$ odd, 
viz. $N = 2K+1$ with $K\in\Nset$.

 In the next section we extend the necessary condition for optimality of \cite{KieJSP} to 
the larger class of cluster problems mentioned above; we state several examples explicitly.
 In section 3 we report the outcome of our test run on the Lennard-Jones cluster data of \cite{CCD}, \cite{Doye}, \cite{Shao},
and acknowledge the comments on the results received after a first version of this paper became publicly available as preprint.
 Section 4 reports the results of the test run on the short data list in \cite{JWHR} for the 
(H$_2$O)$_N$ cluster model featuring monomers with TIP5P five-site potential. 
 We conclude in section 5.

 \vspace{-.5truecm}

\section{A necessary condition for optimality}\vspace{-.15truecm}

We begin by stipulating our notation, and assumptions.

 By $S\in {\mathfrak{S}}$ we denote a \emph{state} of a monomer, a point in a \emph{monomer state space} $\mathfrak{S}$.
 By ${\cal C}^{(N)} :=\{ S_1, ..., S_N\}$ we denote the \emph{configuration} of an $N$-monomer cluster.
 Examples will be given below.

 The following three assumptions characterize the class of cluster models for which we establish a necessary condition for
optimality.
 Namely, first of all, any two monomers in a cluster interact through a symmetric, scalar pair energy function
$V(S,\tilde{S})= V(\tilde{S},S)$ of the state variables of the pair of monomers. 
 Second, the total interaction energy $W({\cal C}^{(N)})$ of an $N$-monomer cluster is the sum over all pairwise 
monomer-monomer interaction energies in the cluster, viz.
\begin{equation}
W({\cal C}^{(N)} ) 
= \sum\!\!\sum_{\hskip-.6truecm 1\leq j< k \leq N} V(S_j,S_k);
\end{equation}
note that this expression for $W$ implies that any internal monomer energy (associated, e.g., with bending, twisting,
stretching, or such) is ignored as irrelevant for cluster formation, and more-than-two-monomer interactions are ignored as
well.
Third, $V$ is such that for any $N$ there exists a globally minimizing configuration ${\cal C}^{(N)}_{\rm min}$ of $W$.

 Before moving on we pause to list
three explicit examples of pair interactions $V$ that satisfy our general assumptions, and which are used in molecular cluster studies.
In all these examples the monomers have a rigid shape, and the pair interaction between two monomers depends only on their
state variables that determine their location and orientation.

 Example {\bf A} represents a type known as single-site potential, used in clusters models when the monomers are neutral spherical atoms.
 Example {\bf B} represents a type known as multi-site potential, used e.g. for clusters of electrically neutral
monomers with polyhedral geometry and electric multi-pole moments.
Example {\bf C} represents a type known as ``chiropole potential, a ``coarse-grained single-site chiral potential,'' 
used to model clusters of a certain bacteriophage.
\medskip

{\bf A}) In the Lennard-Jones cluster problem, $\mathfrak{S} = \Rset^3$, and $S = \rV\in\Rset^3$ is a position vector 
in $3D$ Euclidean space, representing e.g. the location of the center of mass of a single atom. 
 The pair interaction reads (in suitable units of length and energy)
\begin{equation}
V(\rV,\tilde{\rV}) := \frac{1}{r^{12}} - \frac{1}{r^6},
\end{equation}
with $r=|\rV-\tilde{\rV}| \in (0,\infty)$ the distance between the position vectors of the two atoms.

{\bf B}) In the TIP5P model of H$_2$O clusters one has $\mathfrak{S} = \Rset^3\times \Sset^2\times \Sset^1$, 
where the first factor accounts for the position vector $\rV$ of the oxygen atom, and the remaining two factors
for the three Euler angles $(\Theta,\Psi,\Phi)$ that fix the orientation of the rigid tetrahedral charge 
distribution of the monomer, w.r.t. a reference frame.
 The pair interaction is conveniently written (in suitable units of charge, length and energy) as
\begin{equation}
V(S,\tilde{S}) :=
 \frac{1}{r^{12}} - \frac{1}{r^6} 
+
\sum_{a \in\{1,...,4\}} \sum_{b \in\{1,...,4\}} \frac{q_a q_b}{r_{ab}},
\end{equation}
where $r\in (0,\infty)$ is the distance between the position vectors of the two oxygen atoms, $r_{ab}\in (0,\infty)$
is the distance between the $a$-th charge on one H$_2$O monomer and the $b$-th charge on the other, with
$q_a$ and $q_b$ the respective charges on the pertinent monomers.
 Note that once the states $S$ and $\tilde{S}$ are given, the distances $r$ and $r_{ab}$ are determined. 
 The charges of a monomer satisfy $|q_a| = q$ and $\sum_{a}q_a=0$.

{\bf C}) The model of $f\!d$ bacteriophage clusters proposed in \cite{SOKMW} works with $\mathfrak{S} = \Rset^3\times \Sset^2$,
where the first factor accounts for the position vector $\rV$ of the center of a polarized rod of length $2L>0$, the second
factor for the orientation (unit) vector $\muV$ of the chiral poles of that rod, w.r.t. a reference frame. 
 The pair interaction is conveniently written (in suitable units of length and energy) as
\begin{equation}
V(S,\tilde{S}) :=
 \frac{1}{d^{12}} - \frac{1}{d^6}
-
\frac{\Lambda}{r^4}\bigl(\muV\cdot{\tilde{\muV}}\; r\cos\alpha +(\muV\times{\tilde{\muV}})\cdot(\rV-\tilde{\rV})\sin\alpha \bigr),
\end{equation}
where $r\in (0,\infty)$ is the distance between the position vectors of the two rod centers, 
$d\in(0,r]$ is the distance between the two rods (in the usual mathematical sense of distance between two finite
non-intersecting line segments in $\Rset^3$ of length $2L$, a function of only the state variables of the two rods and the parameter $L$),
$\alpha\in [-\frac{\pi}{2},\frac{\pi}{2}]$ is the chiral twist angle of the rod, and $\Lambda>0$ is a coupling constant.
\medskip

This ends our listing of examples. 
 Incidentally, the Lennard-Jones interaction plays a role in all three of them.
We note that the interaction in example {\bf B} reduces to the interaction of example {\bf A} when all $q_a=0$.
Similarly, {\bf C} reduces to {\bf A} in the double limit $\Lambda\to 0$ and $L\to 0$.
 Yet {\bf B} and {\bf C} are not merely small perturbations of {\bf A} but computationally much more challenging than {\bf A}.
 Not surprisingly, the list of putative $N$-monomer cluster optimizers is much longer for example {\bf A} 
than for examples {\bf B} or {\bf C}. \medskip

 Next we define the average pair energy of an $N$-monomer configuration by
\begin{equation}\label{eq:AveV}
\langle V\rangle({\cal C}^{(N)}) := \frac{2}{N(N-1)} \sum\!\!\sum_{\hskip-.6truecm 1\leq j< k \leq N} V(S_j,S_k),
\end{equation}
and the optimal average pair energy as
\begin{equation}
v(N):= \inf_{\cC^{(N)}} \langle V\rangle \big(\cC^{(N)}\big).
\end{equation}
 We here are only interested in models for which the infimum is actually achieved by a minimizing configuration $\cC^{(N)}_{\rm min}$;
whence
\begin{equation}
v(N) = \langle V\rangle \big(\cC^{(N)}_{\rm min}\big).
\end{equation}

This concludes our list of definitions, assumptions, and examples.
\smallskip

\subsection{A graph-theoretical identity}

 Monomers can have quite complicated structures, yet for the (energy) identity we establish next this does not matter. 
 All that matters is that there are $N$ monomers, and that only pairwise interactions are involved that 
satisfy Newton's ``actio equals re-actio'' principle, encoded in the symmetry $V(S,\tilde{S}) = V(\tilde{S},S)$.

 Thus we identify an $N$-monomer cluster, as defined above, with a complete, labeled, non-directed $N$-graph $K_N$.
 The vertices, labeled with $1,2,...,N$, represent the monomers.
 Every edge of $K_N$ represents a pairwise interaction between the monomers that are
represented by the two vertices connected by the edge.
 The edge that connects the $j$-th and $k$-th vertex is assigned the weight $V(S_j,S_k)$; 
note that the edges are without directions, in accordance with the symmetry $V(S_k,S_j) = V(S_j,S_k)$.
 Thus, graph-theoretically the total energy is simply the sum of the edge-weighting function $V$ over all edges.

 Now consider the $\ell$-th subgraph obtained from $K_N$ by removing from $K_N$ the $\ell$-th vertex 
and, hence, with it all edges emanating from it.
 If we now sum the edge-weighting function over all edges of that $\ell$-th subgraph, and then sum that result over all
$\ell$ from 1 to $N$, i.e. sum over all vertices of $K_N$, then all edges appear in the total sum an equal number of times, more than once
yet less than $N$ times because each edge is missing twice from the $N$-fold sum over the vertices, namely the $(j,k)$-edge
is missing when $j=\ell$ and when $k=\ell$. 
 Thus we overcount the total energy by a factor $(N-2)$, i.e. we get
\begin{equation}\label{eq:Eid}
\sum_{1\leq \ell\leq N}
\sum\!\!\sum_{\hskip-.5truecm \begin{subarray}{c}1\leq j< k \leq N\\ \!\! k\neq \ell\neq j\end{subarray}}
 V(S_j,S_k)
=
(N-2) \sum\!\!\sum_{\hskip-.6truecm 1\leq j< k \leq N} V(S_j,S_k).
\end{equation}

 Finally, dividing the graph-theoretical identity (\ref{eq:Eid}) by $N(N-1)(N-2)/2$ and recalling the definition (\ref{eq:AveV})
of the average pair energy of an $N$-monomer configuration, the graph-theoretical identity becomes
\begin{equation}\label{eq:AveVid}
\langle V\rangle({\cal C}^{(N)}) = \frac1N \sum_{1\leq \ell\leq N} \langle V\rangle({\cal C}^{(N)}\backslash\{S_\ell\}).
\end{equation}
\medskip

 Identity (\ref{eq:AveVid}) can be summarized verbally as follows:
\bigskip

\centerline{\emph{The average pair energy of an $N$-monomer cluster equals the arithmetic}}

\centerline{\emph{mean over the average pair energies of all its $N\!-\!1$-monomer sub-clusters}.}

\subsection{Monotonicity of the optimal average pair energy}

 Since identity (\ref{eq:AveVid}) is true for all $\cC^{(N)}$, it holds in particular for the minimizing configuration $\cC^{(N)}_{\rm min}$. 
 But then, since each $\cC^{(N)}_{\rm min}\backslash\{S_\ell\}$ is an $N-1$-monomer configuration that is not
inevitably an energy-minimizing $N-1$-monomer configuration, by replacing each $\cC^{(N)}_{\rm min}\backslash\{S_\ell\}$ 
with the energy-minimizing $N-1$-monomer configuration $\cC^{(N-1)}_{\rm min}$, for all $N>2$ we obtain  
(cf. \cite{KieJSP} for when monomers were just point particles)
\begin{equation}
\langle V\rangle \big(\cC^{(N)}_{\rm min}\big) = 
\frac1N \sum_{\ell=1}^N  \langle V\rangle \big(\cC^{(N)}_{\rm min}\backslash\{S_\ell\}\big) 
\geq \frac1N \sum_{\ell =1}^N  \langle V\rangle \big(\cC^{(N-1)}_{\rm min}\big) 
= \langle V\rangle \big(\cC^{(N-1)}_{\rm min}\big).
\end{equation}
  In short, we have shown that for all $N\geq 2$,
\begin{equation}\label{monoLAW}
\centerline{\boxed{\phantom{\Big[}v(N+1) \geq v(N) \phantom{\Big]}}}\hspace{-1truecm}
\end{equation}
  This a-priori inequality is satisfied by any list $N\mapsto v(N)$ of the average pair energies of 
truly (energy-)optimal $N$-monomer clusters with only pairwise interactions.

\subsection{A test for optimality}

 Since there is no sure-fire algorithm that finds an optimizing $N$-monomer cluster configuration in polynomial time,
the publicly available data lists do not inevitably feature truly optimal clusters, except for small enough $N$.
 Checking lists of putative optimizers for whether they obey this monotonicity law may reveal
data points that fail the test, whence identifying $N$-body cluster energies that are not optimal.

To fail the monotonicity test means the following. Suppose in some computer-experimentally determined
list of putatively optimal $N$-body configurations one finds for a certain $N_*$ that 
$\langle V\rangle \big(\cC^{(N_*-1)}\big) > \langle V\rangle \big(\cC^{(N_*)}\big)$. 
 In that case, and provided that there is no error in the transcription of the data,
 one can conclude that the $N_*-1$-monomer cluster is certainly not optimal.
 Moreover, even if $\langle V\rangle \big(\cC^{(N_*-2)}\big) \leq \langle V\rangle \big(\cC^{(N_*-1)}\big)$,
having discovered that the $N_*-1$-monomer cluster is not optimal, also the $N_*-2$-monomer cluster 
is in limbo (and so on); namely, each $N_*-k$-monomer configurations for which 
$\langle V\rangle \big(\cC^{(N_*-k)}\big) > \langle V\rangle \big(\cC^{(N_*)}\big)$, with $k\geq 1$,
is not optimal.

 Note that for each non-optimal configuration with $N = N_*-k$, 
the difference $\langle V\rangle \big(\cC^{(N_*-k)}\big) - \langle V\rangle \big(\cC^{(N_*)}\big)$
is a lower estimate for the amount by which the average pair energy of $\cC^{(N_*-k)}$ overshoots the optimal value.

\subsection{Bound on missing data}

{In response to a question by the referee, we here add that the monotonicity (\ref{monoLAW}) of the map $N\mapsto v(N)$ 
for optimal average pair energies implies that any value $\langle V\rangle \big(\cC^{(N)}\big)$ 
computed with a putatively optimal $\cC^{(N)}$ is an upper bound on all $v(\tilde{N})$ with $\tilde{N}<N$.
 And so, even if no putatively optimizing configurations $\cC^{(\tilde{N})}$ have yet been computed for certain $\tilde{N}<N$,
 (\ref{monoLAW}) yields some information about the pertinent missing optimal energies 
in such lists of putatively optimal energies that have gaps.}

\vspace{-.5truecm}
\section{Test results for Lennard-Jones data lists}\vspace{-.15truecm}

 The Lennard-Jones data available at the Cambridge Cluster Energy Landscape website \cite{CCD} through a link that
takes you to \cite{Doye} were downloaded and inspected on January 23, 2023.
 Another link there takes you to the Nankai University's Chemoinformatics webpage \cite{Shao} that yields further Lennard-Jones data,
which were downloaded and inspected on April 10, 2023. 
 In total this produced a list of 1609 consecutive data points with $2\leq N\leq 1610$ pooled from 17 different sources,
stated at \cite{Doye}; for instance, the first 110 data appeared in \cite{WD}.
 From those reported configurational energies that we denote by $E^x(N)$, where the superscript ${}^x$ indicates that these
are ``experimental'' data (obtained through what justifiably are called computer experiments),
the average pair energy $v^x(N)$ of each putatively optimal $N$-body cluster was computed 
and then checked for whether the data satisfy the inequality $v^x(N+1)\geq v^x(N)$.
 Remarkably, most data passed this test, but not all. 

 The monotonicity test found that for $N= 447$ the putatively optimal configurational energy is not optimal.
 The ensuing table lists the configurational energies $E^x(N)$ for $N\in\{446,447,448\}$ 
obtained from \cite{CCD},\cite{Doye},\cite{Shao}, and the pertinent average pair energies $v^x(N)$.
 These data lists have been computed with a scaled Lennard-Jones pair potential $V(r) := 4\big( \frac{1}{r^{12}} - \frac{1}{r^6}\big)$, 
normalized so that the ground state energy of a single pair equals $-1$.

\begin{center}
\begin{tabular}{||c c c||} 
 \hline
 $N$ & $E^x(N)$ & $v^x(N)$ \\ 
 \hline\hline
 446 & -2985.461112 & -0.03008475953 \\ 
 \hline
 447 & -2292.783729 & -0.02300121115 \\
 \hline
 448 & -3000.081058 & -0.02996245863  \\ 
 \hline
\end{tabular}
\end{center}
Manifestly, $v^x(448)<v^x(447)$.
 Yet $v^x(447) > v^x(446)$, and also $v^x(448) > v^x(446)$.
 Hence, \emph{assuming that the reported energies are correct}
one may conclude that the $N=447$ cluster is not optimal, while the reported $N=446$ cluster remains putatively optimal, 
as do all clusters with $N\neq 447$ up to $N=1609$, for they all passed the monotonicity test.
 We remark that although the $N=1610$ cluster does have a larger average pair energy than the $N=1609$ cluster, in absence
of an $N=1611$ data point we cannot rule out a hiccup for $N=1610$.

 Interestingly enough, though, it's only the reported energy of the $N=447$ cluster that failed the test,  not the 
pertinent configuration!
 After the first preprint version of this note became public on May 19, 2023, Christian M\"uller promptly recomputed the
energy of the $N=447$ particle Lennard-Jones cluster reported at \cite{Shao} and found that only the configurational energy 
reported at \cite{Shao}, hence \cite{Doye}, hence \cite{CCD}, was incorrect. 
 The correct value is $E^x(447) = -2992.783729449165$ and, as also noted by M\"uller, essentially agrees with the value 
$-2992.783729$ reported in the original publication \cite{XJCS}.
 It yields $v^x(447) = -0.03002361262$, which restores the monotonicity of $N\mapsto v^x(N)$ in the so-corrected Lennard-Jones
 cluster data list.
 Three days later also Carlos Barr\'on Romero communicated to me that he found a $N=447$ cluster configuration with total enery 
equal to $-2992.783729$; 
subsequently he also pointed out the acknowledgement in \cite{ShaoXiangCai} concerning \cite{XJCS}.
 Soon after, Xueguang Shao confirmed that the incorrect value at the website \cite{Shao} is the result of a misprint.

 And so, while our test revealed that the putatively optimal cluster energy for $N=447$ was not optimal, this turned out to be
a curious misprint in the energy data list at \cite{Shao}, from where it migrated to \cite{CCD}, \cite{Doye}. 
At the end of the day it did not imply that the reported $N=447$ configuration itself was non-optimal. 
 All the same, by identifying an incorrect energy data point the test proved useful yet again.

\vspace{-.5truecm}
\section{Test results for a TIP5P data list}\vspace{-.15truecm}

 To have a test of a cluster list for more complicated pair interactions we inspected the list of 
(H$_2$O)$_N$ clusters, $N\in\{2,...,21\}$, published in \cite{JWHR} that has been computed with the TIP5P model \cite{MahoJorg};
see example {\bf B} above.
 We have downloaded it from \cite{CCD} on May 20, 2023.
 Given the more complicated nature of the pair interaction with its mix of Lennard-Jones
and Coulomb terms that determine the location and orientation of the tetrahedral monomers, the list is very
short, containing only 20 different cluster data.
  Not surprisingly, this short list passed the monotonicity test.
 While this does not imply that the listed configurations are true optimizers, by Bayesian probability it has
increased the odds that they actually are.

\vspace{-.5truecm}
\section{Conclusions}\vspace{-.25truecm}

 It is safe to assume that mysterious misprints in computer-generated outputs are exceptionally rare, 
and that a failure of the monotonicity test for $N\mapsto v^x(N)$ normally reveals a non-optimal configuration,
as it did in \cite{KieJSP}.
 Even then it will not reveal a better configuration.
 Yet, once a listed configuration has been identified as non-optimal because its actual average pair energy fails the monotonicity test, 
the existence of a lower-energy configuration is established, and the experts are then called upon to find 
a putatively optimal cluster that passes the monotonicity test.

 In the absence of algorithms that deliver a truly optimal configuration in polynomial time, the simple
monotonicity test described in this paper will eliminate from the list of putative optimizers at least
all those configurations that fail the test.
 It should be easy enough to implement this test directly into search algorithms that produce the computer-generated lists
of putatively optimal data, thereby enhancing the odds that the published putative optimizers are in fact optimizers.

{In this vein, I am pleased to add that Carlos Barr\'on Romero informed me on May 26, 2023 that, 
prompted by the original preprint of this note, he ran the monotonicity test on his list of Lennard-Jones and Morse 
clusters available at \cite{BRb}, which goes up to $N=2063$, and that all his data passed the test.}

{I close by emphasizing one last time that the monotonicity test described in this paper applies to all cluster models
in which the total cluster energy is given by a sum of pairwise energies, thus neglecting energetic contributions from
individual monomer energies (associated with stretching, bending, twisting, etc. of a monomer), and neglecting energetic
contributions of more than two monomers that do not decompose into a sum of pairwise energies. 
 Testing Lennard-Jones clusters is just one example of many conceivable applications, chosen here because of the benchmark status 
enjoyed by the Lennard-Jones cluster problem for which data lists with many hundreds of putative optimizers are publicly available.
 We already mentioned as examples of models with more complicated pair interactions between monomers
the TIP$n$P water cluster models (see example {\bf B} and section 4), 
the model of $f\!d$ bacteriophage clusters proposed in \cite{SOKMW} (see example {\bf C}),
and a model of virus capsids \cite{FCW} made of protein clusters that are interacting with 
the Paramonov--Yaliraki pair potential, a coarse-grained generalization of
the Lennard-Jones potential for ellipsoids \cite{PY}.
 All those model monomers are rigid, but mathematically this is not a requirement for the monotonicity (\ref{monoLAW}) to hold. 
 In principle one could also work with monomers that consist of two or more connected rigid parts that can move partly independently
of each other without involving a change of monomer energy for that; one may think of a hinge, or a short chain of linked loops (ignoring
the effect of gravity on the movable parts). 
 As long as one equips those monomers with purely pairwise interactions that satisfy the assumptions stated in section 2, the
monotonicity (\ref{monoLAW}) will hold.
 Whether any such model could be chemically interesting is a question that lies beyond the expertise of the author to answer.}

\newpage

\noindent
\textbf{ACKNOWLEDGEMENT}:
The monotonicity test of the Lennard-Jones data list downloaded from \cite{CCD}, \cite{Doye}, \cite{Shao} earlier this year
was run by Rutgers undergraduate Annie Wei, who I thank for her early interest in this project.
I thank Professor Christian M\"uller for communicating to me his recomputation of the configurational energy of the reported 
$N=447$ Lennard-Jones cluster configuration, and for thus identifying the monotonicity failure as a result of an unusual misprint 
in the public data lists. I thank Professor Xueguang Shao for confirming that the data point in question is a misprint.
 My thanks go also to Professor Carlos Barr\'on Romero for communicating his $N=447$ Lennard-Jones
cluster energy, for pointing out \cite{ShaoXiangCai} and \cite{BRb}, and for informing me that his cluster data 
passed the monotonicity test of section 2.3.
I am indebted to Professor David Wales for his kind comments, and in particular for suggesting to inquire into whether
models of water clusters can be subjected to the monotonicty test; it made me realize that the reasoning given in \cite{KieJSP}
was not restricted to the point particle models with pair interactions envisioned in \cite{KieJSP}.
I also thank him for remarking that various bio matter models may be susceptible to the monotonicity test, in particular models of
virus capsids as protein clusters.
Last not least I thank the referee for the constructive comments and questions, and for pointing out refs. \cite{SOKMW},
\cite{PY}, and \cite{FCW}.

\vspace{-.5truecm}

\end{document}